\documentstyle[aps,epsfig,twocolumn]{revtex}

\begin{document}

\twocolumn[\hsize\textwidth\columnwidth\hsize\csname@twocolumnfalse\endcsname

\title{Universal c-axis conductivity of high-$T_c$ oxides 
in the superconducting state}

\author{T. Xiang}

\address{Institute of Theoretical Physics, Academia Sinica, P.O.Box 2735,
Beijing 100080, People's Republic of China}

\author{W. N. Hardy}
\address{Department of Physics and Astronomy, University of British Columbia, 
6224 Agricultural Road, Vancouver, British Columbia, V6T 1Z1, Canada}

\date{\today}

\maketitle
\begin{abstract}
The anisotropy in the temperature dependence of the in-plane and c-axis 
conductivities of high-$T_c$ cuprates in the superconducting state is 
shown to be consistent with a strong in-plane momentum dependence of both the 
quasiparticle scattering rate and the interlayer hopping integral. 
Applying the cold spot scattering model recently proposed by Ioffe and Millis 
to the superconducting state, we find that the c-axis conductivity 
varies approximately as $T^3$ in 
an intermediate temperature regime, in good agreement with the experimental 
result for optimally doped YBa$_2$Cu$_3$O$_{7-x}$ and ${\rm Bi_2Sr_2CaCu_2O_{8+x}}$. 
\end{abstract}

]

%\begin{multicols}{2}

Microwave surface impedance measurements have provided important information
on the pairing symmetry\cite{Hardy} and quasiparticle relaxation\cite{Bonn}
in the superconducting state of high-$T_c$ oxides. For YBa$_2$Cu$_3$O$_{7-x}$
(YBCO) (and similarly for other high-$T_c$ compounds\cite{Lee}), the
in-plane microwave conductivity $\sigma _{ab}$ exhibits a large peak
centered at approximately 25K in the superconducting state \cite
{Zhang94,Hosseini98,Hosseini}. This peak structure of $\sigma _{ab}$ is due
to a competition between the quasiparticles life time and the normal fluid
density. From $T_c$ down to 25K the quasiparticle life time increases much
more rapidly than the decrease of normal fluid density, causing $\sigma
_{ab} $ to rise with decreasing temperature. At low temperature, the
quasiparticle life time reaches a limit and increases very slowly but the
normal fluid density continues to fall, $\sigma_{ab}$ therefore falls with
decreasing temperature. Along the c-axis, the conductivity behaves very
differently \cite{Hosseini98}. In contrast to its in-plane counterpart, $
\sigma _c$ falls below $T_c$ and does not show a conductivity peak. In
optimally doped YBCO, $\sigma _c$ rises slightly below 20K. But in Ba$_2$Sr$
_2$CaCu$_2$O$_8$ (BSCCO)\cite{Latyshev,Broun}, 
La$_{2-x}$Sr$_x$CuO$_4$\cite{Kim}, Tl$_2$Ba$
_2 $CuO$_6$ and Tl$_2$Ba$_2$CaCu$_2$O$_8$\cite{Dulic}, $\sigma _c$ drops
continuously from $T_c$ down to very low temperature, and no upturn appears.

The different temperature dependences of $\sigma _{ab}$ and $\sigma _c$ are
not what one might expect within the conventional theory of anisotropic
superconductors. In this paper we present a detailed theoretical analysis
for the c-axis conductivity. We shall show that the decrease of $\sigma _c$
immediately below $T_c$ is explained by the fact that the region near the 
nodes, where the long lived quasiparticles exist, does not enter into the 
c-axis transport because of the anisotropic interlayer hopping integral. 

Let us first consider the behaviour of quasiparticle scattering in high-Tc
materials. An important feature revealed by photoemission measurements is
that the life time of quasiparticles is long along the Brillouin-zone
diagonals and short along other directions on the Fermi surface\cite{Valla} 
in both the normal and supercondcuting states. 
Based on this experimental result and the anisotropic temperature dependence of the
in-plane and c-axis resistivity, Ioffe and Millis (IM)\cite
{Ioffe98,Stojkovic} proposed a cold spot model to account for the normal
state transport data. They assumed that the scattering rate of
quasiparticles contains a large angular dependent part that vanishes
quadratically as the momentum approaches the $(0,0)-(\pi ,\pi )$ line with negligible
frequency and temperature dependence and an isotropic but temperature
dependent part, i.e. $\Gamma _\theta =\Gamma
_0\cos ^22\theta +\tau^{-1}$, where $\theta $ is the angle between the 
in-plane momentum of the electron and the a-axis. 
This type of the scattering rate was also used by Hussey {\it et al.} 
in the analysis of the angular dependent $c$-axis magnetoresistance\cite{Hussey}. 
In Ref. \onlinecite{Ioffe98}, Ioffe and Millis has further assumed that 
$\tau^{-1}$ has the conventional Fermi liquid form 
$\tau_{FL}^{-1}=\frac{T^2}{T_0}+\tau_{imp}^{-1}$. 
With this phenomenological model, they
gave a good explanation for the temperature dependences of several transport
coefficients in the normal state. Van der Marel\cite{Marel99} has recently shown
that this model also provides a good description for both the in-plane and
c-axis optical conductivities in the normal state.

The cold spot scattering rate, as discussed by IM\cite{Ioffe98}, may be
caused by interaction of electrons with nearly singular $d_{x^2-y^2}$
pairing fluctuations. In the superconducting state, as the $d_{x^2-y^2}$
-wave channel scattering is enhanced, the assumption of cold spot scattering
made by IM is strengthened. This is indeed consistent with the 
recent photoemission data measured by Valla {\it et al} \cite{Valla}. 
Thus a detailed comparison between theoretical
calculations and experimental measurements for the transport coefficients in
the superconducting state provides a crucial test for the cold spot model. 

In high-Tc superconductors, the electronic structure is highly anisotropic.
In particular, the interlayer hopping integral depends strongly on the in
plane momentum of electrons. For tetragonal compounds, the c-axis hopping
integral is shown to have the form
\cite{Novikov,Andersen,Xiang96,Xiang98,Chakravarty} 
\begin{equation}
t_c=-t_{\perp }\cos ^2(2\theta ).  \label{tc1}
\end{equation}
This anisotropic interlayer hopping integral is a basic property of 
high-$T_c$ materials. It results from the hybridization
between the bonding O 2p orbitals and virtual Cu 4s orbitals in each CuO$_2$
plane and holds for all high-$T_c$ cuprates with tetragonal symmetry,
independent of the number of CuO$_2$ layers per unit cell\cite{Xiang96,Xiang98}. 
This form of the $c$-axis hopping integral was first found
in the band structure calculations of 
high-$T_c$ oxides\cite{Novikov,Andersen}. 
However, as shown in Refs. \onlinecite{Xiang96,Xiang98}, it is valid 
irrespective of the approximations used in these calculations. 
For Hg$_2$BaCuO$_4$ or other non-body centered tetragonal
compounds, $t_{\perp }$ is approximately independent of $\theta $. For a
body-centered tetragonal compound, such as La$_{2-x}$Sr$_x$CuO$_4$, $
t_{\perp }$ does depend on $\theta $, but in the vicinity of the gap nodes
it is finite. Since in the superconducting state the physical properties are
mainly determined by the quasiparticle excitations near the gap nodes, for
simplicity we ignore the $\theta $-dependence of $t_{\perp }$ in the
discussion given below.

In YBCO, the CuO planes are dimpled with displacements of O in the $c$
direction. O displacements, together with the CuO chains in YBCO, reduce the
crystal symmetry and introduce a finite hybridization between the $\sigma $
and $\pi $ bands. This hybridization results in a small but finite $t_c$
along zone diagonals which will change the low temperature behavior of the 
electromagnetic response functions. However, at not too low temperatures,
Eq. (\ref{tc1}) is still a good approximation.

The conductivity is determined by the imaginary part of the current-current
correlation function. If vertex corrections are ignored\cite{Hirschfeld93}, 
it can be shown that the conductivity is given by 
\begin{eqnarray}
\sigma _\mu &=&-\frac{\alpha _\mu }\pi \int_{-\infty }^\infty d\omega \frac{
\partial f(\omega )}{\partial \omega }\int_0^{2\pi }\frac{d\theta }{2\pi }
u_\mu ^2(\theta )M(\theta )  \label{sigma} \\
M(\theta ) &=&\frac \pi {\Gamma _\theta }{\rm Re}\frac{\left( \omega
+i\Gamma _\theta \right) ^3-\omega \Delta _0^2\cos ^22\theta }{\left[ \left(
\omega +i\Gamma _\theta \right) ^2-\Delta _0^2\cos ^22\theta \right] ^{3/2}},
\end{eqnarray}
where $u_{ab}(\theta )=1$, $u_c(\theta )=\cos ^2(2\theta )$, $\alpha
_{ab}=e^2v_F^2N(0)/4$, $\alpha _c=e^2t_{\perp }^2N(0)/4$, $N(0)$ is the
density of states of electrons at the Fermi level, and $f(\omega )$ is the
Fermi function. In obtaining the above equation, the retarded Green's
function of the electron, $G_{ret}(k,\omega ),$ is assumed to be 
\begin{equation}
G_{ret}(k,\omega )=\frac 1{\omega -\xi _k\tau _3-\Delta _\theta \tau
_1+i\Gamma _\theta },
\end{equation}
where $\xi _k=\varepsilon _{ab}(k)-t_{\perp }\cos k_zu_c(\theta )$ is the
energy dispersion of the electron, $\Delta _\theta =\Delta _0\cos 2\theta $
is the d-wave gap parameter, $\Gamma _\theta $ is the quasiparticle
scattering rate, $\tau_1$ and $\tau_3$ are the Pauli matrices. 

In the normal state, $\Delta _0=0$, and Eq. (\ref{sigma}) becomes 
\begin{equation}
\sigma _\mu =\alpha _\mu \int_0^{2\pi }\frac{d\theta }{2\pi }\frac{u_\mu
^2(\theta )}{\Gamma _\theta }.  \label{normal}
\end{equation}
If the scattering is isotropic, i.e. $\Gamma _\theta =\Gamma (T)$
independent of $\theta $, then $\sigma _{ab}$ and $\sigma _c$ should have
the same temperature dependence, in contradiction with experiments. However,
if $\Gamma _\theta =\Gamma _0\cos ^22\theta +\tau ^{-1}(T)$, then 
\begin{eqnarray}
\sigma _{ab} &=&\frac{\alpha_{ab}}{\sqrt{\tau ^{-1}(\Gamma _0+\tau ^{-1})}},
\label{sigma_ab_normal} \\
\sigma _c &=&\frac{\alpha _c}{2\Gamma _0}\left( 1-\frac{2\tau ^{-1}}{\Gamma
_0}+\frac{2\tau ^{-1}}{\Gamma _0}\sqrt{\frac{\tau ^{-1}}{\Gamma _0+\tau ^{-1}
}}\right) .
\end{eqnarray}
When $\Gamma _0\gg \tau ^{-1}$, $\sigma _{ab}$ is proportional to $\sqrt{
\tau }$ not $\tau $. This is the result first obtained by IM with a Boltzman
equation analysis. If $\tau ^{-1}$varies quadratically with $T$ as in
conventional Fermi liquid theory, the resistivity, i.e. $\sigma _{ab}^{-1}$,
varies linearly with $T$. This provides a phenomenological account for the
linear resistivity of optimal doped cuprates. $\sigma _c$ depends on two
parameters, $\alpha _c/\Gamma _0$ and $\tau \Gamma _0$. In the limit $\Gamma
_0\gg \tau ^{-1}$, $\sigma _c\ $ depends very weakly on $T$ and extrapolates
to a finite value $\alpha _c/\left( 2\Gamma _0\right) $ at $T=0K$, in
qualitative agreement with the experimental data. These results indicate
that the simple cold spot model captures the key features of high-$T_c$
transport properties in the normal state, although its microscopic 
mechanism is still unclear.

In the superconducting state, Eq. (\ref{sigma}) cannot be integrated out
analytically. However, in the temperature regime $T_c>T\gg $ $\tau
_0^{-1}$, where $\tau_0 = \langle \Gamma^{-1}
_\theta \rangle $ is the thermal average of $\Gamma^{-1}_\theta $, the leading
order approximation in $\Gamma _\theta $ is valid and the conductivity is
given by\cite{Hirschfeld93} 
\begin{equation}
\sigma _\mu \approx -\alpha _\mu \int_{-\infty }^\infty d\omega \frac{
\partial f(\omega )}{\partial \omega }\int_0^{2\pi }\frac{d\theta }{2\pi }
\frac{u_\mu ^2(\theta )}{\Gamma _\theta }{\rm Re}\frac{\left| \omega \right| 
}{\sqrt{\omega ^2-\Delta _\theta ^2}}.  \label{super}
\end{equation}
$\tau _0^{-1}$ can be estimated from the experimental data of the in-plane
microwave conductivity $\sigma _{ab}$ and the normal fluid density with the
generalized Drude formula\cite{Bonn}: $\sigma _{ab}\sim n_{ab}\tau _0$. For
optimally doped YBCO\cite{Hosseini}, $\tau _0^{-1}$ is  less than $1K$ at
low temperatures and increases with increasing temperatures. At $60K$, $\tau
_0^{-1}$ is about $6K$. Close to $T_c$, $\tau _0^{-1}$ becomes larger but
still much less than the temperature. This means that the leading order
approximation in $\Gamma _\theta $ is valid in nearly the whole temperature
range in which the experimental measurements have been done so far, at least
for optimally doped YBCO.

If $\Gamma _\theta =\Gamma (T)=\tau ^{-1}(T)$ does not depend on $\theta $,
Eq. (\ref{super}) can be simplified to 
\begin{equation}
\sigma _\mu =\frac{e^2n_\mu (T)\tau }{2m},  \label{drude}
\end{equation}
where $n_\mu (T)$ is the normal fluid density which decreases with
decreasing temperatures. This is nothing but the generalized Drude formula
which was first used by Bonn {\it et al. }in their data analysis for the
in-plane microwave conductivity in the superconducting state\cite
{Hirschfeld93,Bonn93}. From Eq. (\ref{drude}), it is easy to show that the
ratio of the in- and out-of-plane conductivities $\sigma _{ab}/\sigma _c$ is
proportional to the ratio $n_{ab}/n_c$, i.e. $\sigma _{ab}/\sigma _c=$ $
n_{ab}/n_c$. However, this does not agree with experiments, even
qualitatively. It implies that the scattering rate must be anisotropic, as
mentioned previously.

In the cold spot model, $\Gamma _\theta =\Gamma _0\cos ^22\theta +\tau
^{-1}(T)$ , $\sigma _{ab}$ and $\sigma _c$ behave very differently. Eq. (\ref
{super}) now can be approximately written as 
\begin{eqnarray}
\sigma _a &\approx &-\frac{T\tau \alpha _a}{\Delta _0}\int_{-\infty }^\infty
dx\frac{\partial f(xT)}{\partial x}\frac{\left| x\right| }{\sqrt{1+T^2\Gamma
_0\tau x^2/\Delta _0^2}}, \\
\sigma _c &\approx &\frac{9\alpha _c\zeta [3]T^3}{2\Gamma _0\Delta _0^3}-
\frac{\left( 2\ln 2\right) T\alpha _c}{\tau \Gamma _0^2\Delta _0}+\frac{
\alpha _c\sigma _a}{\alpha _a\tau ^2\Gamma _0^2},
\end{eqnarray}
where $\zeta (3)=1.202$. In the high temperature limit $\Gamma _0\tau
T^2/\Delta _0^2\gg 1$, 
\begin{eqnarray}
\sigma _{ab} &\approx &\alpha _{ab}\sqrt{\frac \tau {\Gamma _0}},
\label{inplane} \\
\sigma _c &\approx &\frac{9\alpha _c\zeta (3)T^3}{2\Gamma _0\Delta _0^3}
= 9\zeta (3) \sigma_{n,c}(0K) \frac{T^3}{\Delta_0^3}, 
\label{cubic}
\end{eqnarray}
where $\sigma_{n,c}(0K) = \alpha_c/2\Gamma_0$ is the extrapolated 
normal state c-axis conductivity at $0K$. 
These equations reveal a few interesting properties of the conductivities.
Firstly, $\sigma _{ab}$ is proportional to $\sqrt{\tau }$ and does not
depend explicitly on $\Delta _0$. This $\sqrt{\tau }$ dependence of $\sigma
_{ab}$ is an extension of Eq. (\ref
{sigma_ab_normal}) in the superconducting state, 
which means that $\sigma _{ab}$ (excluding the
fluctuation peak at $T_c$) will change smoothly across $T_c$ since $\tau $ 
changes continuously at $T_c$. The
temperature dependence of $\tau $ in the superconducting state is unknown.
But phenomenologically it can be determined from the measured in-plane
conductivity via Eq. (\ref{inplane}). Secondly, $\sigma _c$ decreases
monotonically with decreasing temperature and behaves approximately as $T^3$
in the above temperature regime ($\Delta _0$ depends very weakly on
temperature except close to $T_c$). Furthermore $\sigma _c$ does not depend
on $\tau $, which means that this $T^3$ behavior is {\it universal} and
independent of the impurity scattering provided it is sufficiently weak that
the coherent interlayer tunneling dominates. Furthermore, there is 
no free adjustable parameters in Eq. (\ref{cubic}) since both 
$\sigma_{n,c}(0K)$ and $\Delta_0$ can be determined directly from 
experiments. This therefore provides a good
opportunity to test the cold spot scattering model by comparison with
experiments.

In the low temperature limit $\Gamma _0\tau T^2/\Delta _0^2\ll 1$, Eq. (\ref
{super}) leads to the following results 
\begin{eqnarray}
\sigma _{ab} &\approx &\frac{(2\ln 2)T\tau \alpha_{ab}}{\Delta _0},
\label{a1} \\
\sigma _c &\approx &\frac{675\zeta [5]\alpha _c\tau }4\left( \frac T{\Delta
_0}\right) ^5.  \label{c1}
\end{eqnarray}
In this limit both $\sigma _{ab}$ and $\sigma _c$ do not depend on $\Gamma _0
$. This means that the $\Gamma _0\cos ^22\theta $ term in $\Gamma _\theta $
is not important in this temperature regime. In fact, Eqs. (\ref{a1}) and 
(\ref{c1}) are just the results of $\sigma_{ab}$ and $\sigma_c$ in an
isotropic scattering system as given by Eq. (\ref{drude}) since the normal
fluid densities $n_{ab}$ and $n_c$ behave as $T$ and $T^5$ for tetragonal
high-$T_c$ compounds at low temperatures, respectively
\cite{Xiang96,Xiang98,Panagopoulos,Gaifullin}. In
real materials where impurity scattering is not negligible, as discussed in
Refs. \cite{Xiang96,Xiang98},  this $T^5$ behaviour of the $c$-axis normal
fluid density is fairly unstable and will be replaced by a $T^2$ law at low
temperatures. In this case, the temperature dependence of $\sigma _c$ will
also be changed. 

The condition $\Gamma _0\tau T^2/\Delta _0^2\gg 1$ can be written as $
T/\Delta _0\gg 1/\sqrt{\Gamma _0\tau }$. Since $\tau^{-1}_0 > \tau^{-1}$, 
the condition  $T/\Delta _0\gg 1/\sqrt{\Gamma _0\tau }$ holds if  $
T/\Delta _0\gg 1/\sqrt{\Gamma _0\tau_0 }$ is satisfied. 
If we assume $\Gamma _0\sim $ $0.15eV
$ , which is the value used by IM in their analysis of normal state
transport coefficients\cite{Ioffe98}, and $\tau _0^{-1}\sim 6K
$ as given by the experimental data for YBCO at $60K$, then $1/\sqrt{\Gamma
_0\tau_0 }$ is estimated to be about $0.06$. Thus the condition $\Gamma _0\tau
T^2/\Delta _0^2\gg 1$ holds at least when $T/\Delta _0\gg 0.06$ for YBCO. Since $
\Delta _0\sim 2T_c$ at low temperatures, therefore Eqs. (\ref{inplane}) 
and (\ref{cubic}) are valid when $T/T_c\gg 0.12$ for optimally doped YBCO.

\begin{figure}
\centering\epsfig{file=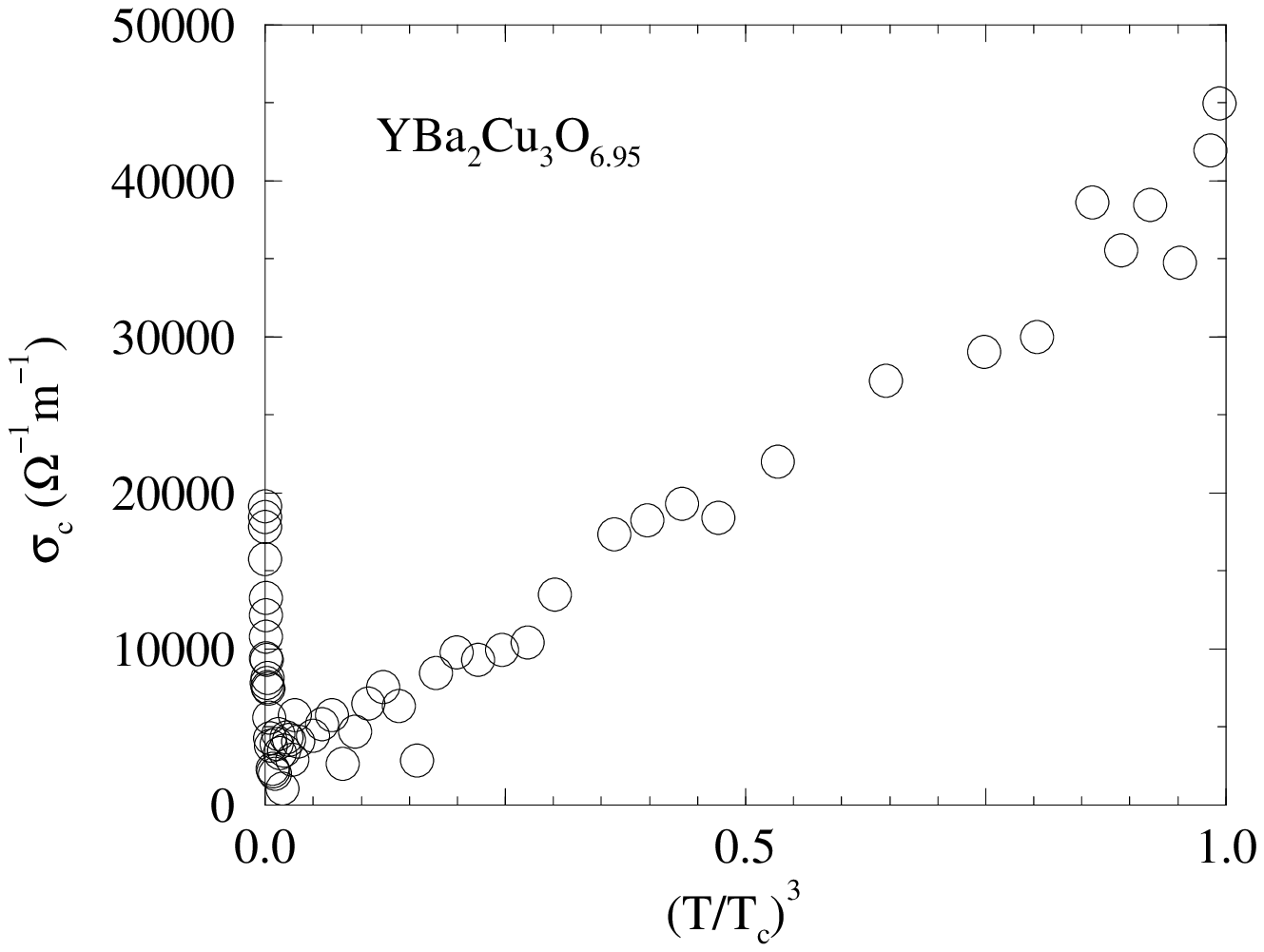,width=\linewidth,clip=,angle=0}
\caption{The c-axis conductivity as a function of $(T/T_c)^3$ for YBa$_2$Cu$
_3$O$_{6.95}$ measured at 22GHz [5]. }
\end{figure}

To compare the above results with experiments, we plot in Figure  1 
the experimental data of $\sigma _c$ at 22GHz 
as a function of $(T/T_c)^3$ for 
${\rm YBa_2Cu_3O}_{6.95}$ \cite{Hosseini98}. From $30K$ up to $T_c$,   
$\sigma _c$ exhibits a $T^3$ behavior 
within experimental error, in agreement with Eq. (\ref{cubic}). 
We have fitted the experimental data with other power laws of 
$T/T_c$ in the same temperature range, but found none of them fit the
experimental data as well as the $T^3$ power law.  For 
this material, there is no experimental data on the 
temperature dependence of $\sigma_c$ in the normal state. 
But just above $T_c$, $\sigma_c \approx 6.3\times 10^4
\Omega^{-1} m^{-1}$ \cite{Hosseini}. Since the normal state $\sigma_c$
depends very weakly on $T$ at low $T$ for 
optimally doped YBCO \cite{Takenaka}, we can therefore
approximately take this value of $\sigma_c$ as the extrapolated normal
state c-axis conductivity at $0K$, i.e. $\sigma_{n,c}(0K) 
\approx 6.3\times 10^4 \Omega^{-1} m^{-1}$. 
Substituting this value into Eq. (\ref{cubic}), we obtain 
$\sigma_c \approx 6.8\times 10^5(T/\Delta_0)^3 \Omega^{-1}m^{-1}$. 
By fitting this theoretical result with 
the corresponding experimental data in Figure 1, 
we find that $\Delta_0/T_c\approx 2.6$. This value of $\Delta_0/T_c$ 
agrees with all other published data for optimally doped 
YBCO within experimental uncertainty. This agreement indicates that 
not only the leading temperature dependence but also the 
absolute values of $\sigma_c$ predicted by 
Eq. (\ref{cubic}) agrees with experiments.

For YBCO, $t_c$ is small but finite at the gap nodes. This has not been
considered in the above discussion and if included, the
temperature dependence of $\sigma _c$ at low temperatures will be changed.
What effect leads to the upturn of $\sigma _c$ at low $T$ remains a puzzle 
to us. The model
presented here is inadequate to address this issue. Perhaps the interplay
between the CuO chains and CuO$_2$ is playing an important role. 
A possible cause
for the upturn of $\sigma _c$ is the proximity effect between CuO chains and
CuO$_2$ planes. However, a detailed analysis for this is too complicated to
carry out at present and at this point we are unable to show explicitly whether 
this effect
alone can lead to the observed upturn in $\sigma _c$. Another possibility is
that this upturn is due to the onset of coherent tunnelling of the $c$-axis
hopping whereas before this upturn the c-axis hopping is incoherent. If
this is the case, the assumption made in this paper would fail. However,
since the dramatic increase in the quasiparticle lifetime 
\cite{Bonn,Hosseini,Krishana} occurs at a temperature much higher t
han the $\sigma_c$ upturn temperature, we believe that this possibility 
is unlikely to be relevant.

\begin{figure}
\centering\epsfig{file=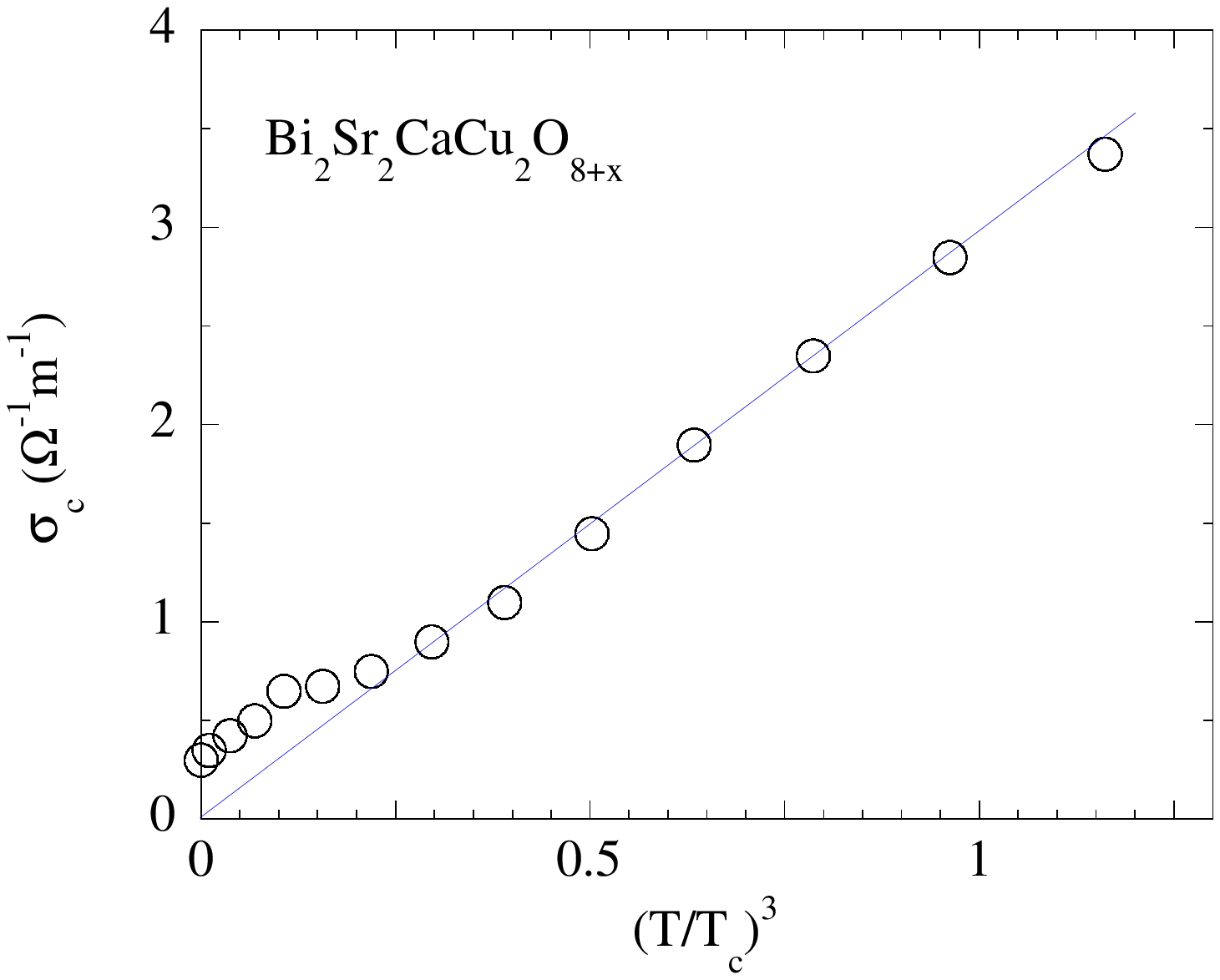,width=\linewidth,clip=,angle=0}
\caption{The c-axis conductivity as a function of $(T/T_c)^3$ for 
${\rm Bi_2Sr_2CaCu_2O_{8+x}}$ ($T_c = 78K$) [7]. The solid line is 
provided as a guide only.}
\end{figure}

Figure 2 shows the $c$-axis quasiparticle conductivity 
obtained by Latyshev {\it et al.} from an intrinsic mesa 
tunneling measurement\cite{Latyshev} for BSCCO. $\sigma_c$ of 
BSCCO also exhibits a $T^3$ behavior in a broad 
temperature range in the superconducting state. However, the 
onset temperature of this $T^3$ term ($\sim 45K$) is higher than that 
for YBCO. This is because BSCCO is very anisotropic and 
disorder effects are stronger than in YBCO. 
The crossover temperature from the impurity dominated limit at low 
temperatures to the intrinsic limit at high temperatures, as  
estimated in Ref. \onlinecite{Latyshev}, is about 30K. 
The value of $\sigma_{n,c}(0K)$ for this material is difficult 
to determine because of the pseudogap effect. Since the opening of 
a pseudogap always reduces the value of $\sigma_{n,c}$ 
in the normal state, we can therefore take the 
value of $\sigma_{n,c}$ just above $T_c$ ($\sim 3\Omega^{-1}m^{-1}$), 
as a lower bound to $\sigma_{n,c}(0K)$. 
The c-axis conductivity measured at a voltage well above the pseudogap 
is nearly temperature independent and higher than the corresponding 
conductivity in the limit $V\rightarrow 0$
(Figure 2 of Ref. \cite{Latyshev}). This conductivity, 
$\sigma_{n,c}(eV >\Delta_0)\sim 8\Omega^{-1}m^{-1}$, sets a upper bound to 
$\sigma_{n,c}(0K)$. Thus $\sigma_{n,c}(0K)$ is between 
$3\Omega^{-1}m^{-1}$ and
$8\Omega^{-1}m^{-1}$. By fitting the experimental data of $\sigma_c$ from 45K to 
$T_c$ with Eq. \ref{cubic}, we find that $\Delta_0/T_c$ is within (2.3, 3.1). 
This range of $\Delta_0/T_c$ is consistent with the 
published data for BSCCO within experimental uncertainty.

In conclusion, we have studied the temperature behavior of microwave
conductivity of high-$T_c$ cuprates in the superconducting state within the
framework of low energy electromagnetic response theory of superconducting
quasiparticles. We found that the c-axis conductivity varies approximately
as $T^3$ and does not depend on $\tau (T)$ in the temperature regime $T_c\gg
T\gg \Delta _0/\sqrt{\Gamma _0\tau }$.
This {\it universal} tempearture dependence of $\sigma_c$ 
agrees quantitatively with the experimental data for YBCO and BSCCO. 
Our study shows that it is important to include the anisotropy of 
the interlayer hopping integral in the analysis of $c$-axis 
transport properties of high-$T_c$ cuprates.

We thank D. Broun, A. J. Millis, C. Panagopoulos, A. J. Schofield 
for useful discussions, and Yu. I. Latyshev for sending us the 
experimental data as shown in Figure 2.  T.X. was supported in part by 
the National Science Fund for Distinguished Young Scholars of 
the National Natural Science Fundation of China.

%\end{multicols}

\end{document}